\documentclass[runningheads]{llncs}
\usepackage[T1]{fontenc}
\usepackage{graphicx}

\usepackage{euscript}
\usepackage{amssymb}
\usepackage{multirow}
\usepackage[misc]{ifsym}
\usepackage[super]{nth}
\usepackage{diagbox}
\usepackage{hhline}
\usepackage{subcaption}
\usepackage[hidelinks]{hyperref}

\usepackage[table]{xcolor}
\usepackage{ulem}

\definecolor{franzcolor}{rgb}{0.1, 0.8, 0.4}

\definecolor{darkocolor}{rgb}{0.1, 0.4, 0.8}

\definecolor{martincolor}{rgb}{0.1, 0.4, 0.8}

\definecolor{redcolor}{rgb}{0.8, 0.0, 0.0}

\definecolor{bluecolor}{rgb}{0.0, 0.0, 0.8}

\definecolor{orangecolor}{rgb}{0.9, 0.5, 0.0}

\newcommand{\boldparagraph}[1]{\paragraph{{\textbf{#1}}}}

\usepackage{xcolor,colortbl}
\usepackage{colortbl}

\newcommand{\lightvalue}{0.92}
\definecolor{lightgray}{rgb}{\lightvalue,\lightvalue,\lightvalue}
\definecolor{lightgreen}{rgb}{\lightvalue,1,\lightvalue}
\definecolor{lightred}{rgb}{1,\lightvalue,\lightvalue}
\definecolor{lightblue}{rgb}{\lightvalue,\lightvalue,1}
\definecolor{lightyellow}{rgb}{1,1,\lightvalue}

\usepackage{arydshln}

\newcommand{\midlightvalue}{0.8}
\definecolor{midlightred}{rgb}{1,\midlightvalue,\midlightvalue}
\definecolor{midlightgreen}{rgb}{\midlightvalue,1,\midlightvalue}
\definecolor{midlightblue}{rgb}{\midlightvalue,\midlightvalue,1}
\definecolor{midlightyellow}{rgb}{1,1,\midlightvalue}
\definecolor{midlightcyan}{rgb}{\midlightvalue,1,1}
\definecolor{midlightmagenta}{rgb}{1,\midlightvalue,1}

\definecolor{midlightmscmr1}{rgb}{0.5,0.5,\midlightvalue}
\definecolor{midlightmscmr2}{rgb}{\midlightvalue,0.5,0.5}
\definecolor{midlightmscmr3}{rgb}{0.5,\midlightvalue,0.5}

\definecolor{todocolor}{rgb}{1.0,0.7,0.4}

\definecolor{notecolor}{rgb}{0.5,1.0,0.5}
\makeatletter
\define@key{todonotes}{note}[]{%
    \setkeys{todonotes}{color=notecolor}}%
\makeatother

\usepackage{dcolumn}        
\newcolumntype{C}{D{.}{.}{2.2}} 

\usepackage{glossaries}

\newacronym{scn}{SCN}{Spatial-ConfigurationNet}
\newacronym{nnunet}{nnU-Net}{no-new-Net}
 
\newacronym{gdloss}{GD loss}{Generalized Dice Loss}
\newacronym{celoss}{CE loss}{Cross-Entropy Loss}

\newacronym[plural=CNNs,firstplural=Convolutional Neural Networks (CNNs)]{cnn}{CNN}{Convolutional Neural Network}
\newacronym[plural=ASMs,firstplural=Active Shape Models (ASM)]{asm}{ASM}{Active Shape Model}
\newacronym[plural=AAMs,firstplural=Active Appearance Models (AAM)]{aam}{AAM}{Active Appearance Model}
\newacronym{cca}{CCA}{connected component analysis}
\newacronym{mas}{MAS}{Multi-Atlas Segmentation}
\newacronym{mcd}{MCD}{Monte-Carlo Dropout}

\newacronym{iid}{i.i.d.}{independent and identically distributed}
\newacronym{tl}{TL}{Transfer Learning}
\newacronym{da}{DA}{Domain Adaptation}
\newacronym{dg}{DG}{Domain Generalization}
\newacronym{dr}{DR}{Domain Randomization}
\newacronym{sfda}{SFDA}{Source-Free Domain Adaptation}
\newacronym{tta}{TTA}{Test-time Adaptation}

\newacronym{sap}{SAP}{Semantics-Aware Augmentation Pipeline}
\newacronym{lsa}{LSA}{Label-Specific Intensity Augmentation}
\newacronym{rna}{RNA}{Random Convolution Network Augmentation}
\newacronym{tim}{TIM}{Target-to-source Intensity Matching}

\newacronym{iamdg}{IAM}{Intensity Augmentation and Matching}

\newacronym{ct}{CT}{Computed Tomography}
\newacronym{mr}{MR}{Magnetic Resonance}
\newacronym{hu}{HU}{Hounsfield Units}

\newacronym{miccai}{MICCAI}{International Conference on Medical Image Computing and Computer-Assisted Intervention}
\newacronym{mmwhs}{MMWHS}{Multi-Modality Whole Heart Segmentation}
\newacronym{flare}{FLARE}{Fast and Low GPU Memory Abdominal Organ Segmentation}
\newacronym{amos}{AMOS}{Multi-Modality Abdominal Multi-Organ Segmentation}
\newacronym{bcv}{BCV}{Multi-Atlas Labeling Beyond the Cranial Vault}
\newacronym{bcvc}{BCV-C}{BCV-Cervix}
\newacronym{bcva}{BCV-A}{BCV-Abdomen}
\newacronym{chaos}{CHAOS}{Combined Healthy Abdominal Organ Segmentation}

\newacronym[plural=ROIs,firstplural=regions of interest (ROI)]{roi}{ROI}{region of interest}
\newacronym[plural=GPUs,firstplural=graphics processing units (GPU)]{gpu}{GPU}{graphics processing unit}

\newacronym{lv}{LV}{left ventricle}
\newacronym{rv}{RV}{right ventricle}
\newacronym{la}{LA}{left atrium}
\newacronym{ra}{RA}{right atrium}
\newacronym{myo}{MYO}{myocardium}
\newacronym{aa}{AA}{ascending aorta}
\newacronym{pa}{PA}{pulmonary artery}

\newacronym{cg}{CG}{central gland}
\newacronym{pz}{PZ}{peripheral zone}

\newacronym{li}{LI}{liver}
\newacronym{lk}{LK}{left kidney}
\newacronym{rk}{RK}{right kidney}
\newacronym{sp}{SP}{spleen}
\newacronym{st}{ST}{stomach}

\newacronym{bmc}{BMC}{Boston Medical Center}
\newacronym{runmc}{RUNMC}{Radboud University Nijmegen Medical Center}

\newacronym{dsc}{DSC}{Dice Similarity Coefficient}
\newacronym{gdsc}{G-DSC}{Generalized Dice Similarity Coefficient}
\newacronym{assd}{ASSD}{Average Symmetric Surface Distance}
\newacronym{hd}{HD}{Hausdorff Distance}

\newacronym{acc}{ACC}{Accuracy}
\newacronym{sen}{SEN}{Sensitivity}
\newacronym{spe}{SPE}{Specificity}
\newacronym{pre}{PRE}{Precision}

\newacronym{rsc}{RSC}{representation self-challenging}

\newacronym[plural=MS-CaRe-CNNs,firstplural=Multi-Sequence Cascading Refinement CNNs (MS-CaRe-CNN)]{mscarecnn}{MS-CaRe-CNN}{Multi-Sequence Cascading Refinement CNN}

\newacronym{lge}{LGE}{late gadolinium enhanced}
\newacronym{t2}{T2}{T2-weighted}
\newacronym{bssfp}{bSSFP}{balanced steady-state free precession}

\newacronym{myosaiq}{MYOSAIQ}{Myocardial Segmentation with Automated Infarct Quantification}
\newacronym{fimh}{FIMH}{International Conference on Functional Imaging and Modeling of the Heart}
\newacronym{emidec}{EMIDEC}{\textit{Automatic Evaluation of Myocardial Infarction from Delayed-Enhancement Cardiac MRI}}

\newacronym{mi}{MI}{myocardial infarction}

\newacronym{mit}{MIT}{myocardial infarct tissue}
\newacronym{mvo}{MVO}{microvascular obstruction}
\newacronym{fmyo}{f-MYO}{full myocardium}
\newacronym{fmit}{f-MIT}{full myocardial infarct tissue}

\newacronym{cc}{CC}{correlation coefficient score}
\newacronym{mae}{MAE}{mean absolute error}
\newacronym{loa}{LOA}{limits of agreement}
\newacronym{crps}{CRPS}{continuous ranked probability score}

\newacronym{psir}{PSIR}{phase sensitive inversion recovery}
\newacronym{mag}{MAG}{magnitude reconstruction}
\newacronym{bmm}{BMM}{boundary mining model}
\newacronym{alm}{ALM}{adversarial learning model}

\newacronym{cdcblock}{CDC block}{convolution-dropout-convolution block}

\newcommand{\lossbase}{L}
\newcommand{\lossdice}{\lossbase_{\text{GD}}}

\newcommand{\imagelettersmall}{x}
\newcommand{\groundtruthlettersmall}{y}

\newcommand{\predictionlettersmall}{p}

\newcommand{\image}{\mathbf{\imagelettersmall}}
\newcommand{\groundtruth}{\mathbf{\groundtruthlettersmall}}

\newcommand{\prediction}{\mathbf{\hat{\groundtruthlettersmall}}}

\newcommand{\predlettersmall}{\mathbf{\hat{\predictionlettersmall}}}
\newcommand{\predstageone}{\predlettersmall_1}
\newcommand{\predstagetwo}{\predlettersmall_2}

\newcommand{\labelpredstageone}{\mathbf{\hat{\groundtruthlettersmall}}_1}
\newcommand{\labelpredstagetwo}{\mathbf{\hat{\groundtruthlettersmall}}_2}

\newcommand{\gtstageone}{\mathbf{\groundtruthlettersmall}_1}
\newcommand{\gtstagetwo}{\mathbf{\groundtruthlettersmall}_2}

\newcommand{\modelweightsbase}{\theta}
\newcommand{\modelweightsstageone}{{\modelweightsbase}_1}
\newcommand{\modelweightsstagetwo}{{\modelweightsbase}_2}

\newcommand{\modelbase}{\mathcal{M}}
\newcommand{\modelstageone}{{\modelbase}_1}
\newcommand{\modelstagetwo}{{\modelbase}_2}

\newcommand{\lossfactorbase}{\lambda}
\newcommand{\lossfactorone}{{\lossfactorbase}_1}
\newcommand{\lossfactortwo}{{\lossfactorbase}_2}

\newcommand{\groupone}{\text{G1}}
\newcommand{\grouptwo}{\text{G2}}
\newcommand{\groupthree}{\text{G3}}

\newcommand{\datagroupone}{\mathcal{D}^{\groupone}}
\newcommand{\datagrouptwo}{\mathcal{D}^{\grouptwo}}
\newcommand{\datagroupthree}{\mathcal{D}^{\groupthree}}

\newcommand{\imagegroupone}{\image^{\groupone}}
\newcommand{\imagegrouptwo}{\image^{\grouptwo}}
\newcommand{\imagegroupthree}{\image^{\groupthree}}

\newcommand{\predstageonegroupone}{\predlettersmall_{1}^{\groupone}}
\newcommand{\predstageonegrouptwo}{\predlettersmall_{1}^{\grouptwo}}
\newcommand{\predstageonegroupthree}{\predlettersmall_{1}^{\groupthree}}

\newcommand{\predstagetwogroupone}{\predlettersmall_{2}^{\groupone}}
\newcommand{\predstagetwogrouptwo}{\predlettersmall_{2}^{\grouptwo}}
\newcommand{\predstagetwogroupthree}{\predlettersmall_{2}^{\groupthree}}

\usepackage{color}

\begin{document}
\title{Multi-Source and Multi-Sequence \\ Myocardial Pathology Segmentation \\ Using a Cascading Refinement CNN}

\titlerunning{Multi-Source and Multi-Sequence Myocardial Pathology Segmentation}

\author{
Franz Thaler\inst{1,2}\orcidID{0000-0002-6589-6560} \and
Darko \v{S}tern\inst{3}\orcidID{0000-0003-3449-5497} \and
Gernot Plank\inst{1}\orcidID{0000-0002-7380-6908} \and
Martin Urschler\inst{4}\orcidID{0000-0001-5792-3971}
}

\authorrunning{F. Thaler et al.}

\institute{
Gottfried Schatz Research Center: Medical Physics and Biophysics, \\Medical University of Graz, Graz, Austria \and
Institute of Computer Graphics and Vision, Graz University of Technology, \\Graz, Austria \and
AVL List GmbH, Graz, Austria \and
Institute for Medical Informatics, Statistics and Documentation, \\Medical University of Graz, Graz, Austria\\
}


\maketitle              
\begin{abstract}

Myocardial infarction (MI) is one of the most prevalent cardiovascular diseases and consequently, a major cause for mortality and morbidity worldwide.
Accurate assessment of myocardial tissue viability for post-MI patients is critical for diagnosis and treatment planning, e.g. allowing surgical revascularization, or to determine the risk of adverse cardiovascular events in the future.
Fine-grained analysis of the myocardium and its surrounding anatomical structures can be performed by combining the information obtained from complementary medical imaging techniques.
In this work, we use late gadolinium enhanced (LGE) magnetic resonance (MR), T2-weighted (T2) MR and balanced steady-state free precession (bSSFP) cine MR in order to semantically segment the left and right ventricle, healthy and scarred myocardial tissue, as well as edema.
To this end, we propose the Multi-Sequence Cascading Refinement CNN (MS-CaRe-CNN), a 2-stage CNN cascade that receives multi-sequence data and generates predictions of the anatomical structures of interest without considering tissue viability at Stage~1.
The prediction of Stage~1 is then further refined in Stage~2, where the model additionally distinguishes myocardial tissue based on viability, i.e. healthy, scarred and edema regions.
Our proposed method is set up as a 5-fold ensemble and semantically segments scar tissue achieving $62.31$\% DSC and $82.65$\% precision, as well as $63.78$\% DSC and $87.69$\% precision for the combined scar and edema region.
These promising results for such small and challenging structures confirm that MS-CaRe-CNN is well-suited to generate semantic segmentations to assess the viability of myocardial tissue, enabling downstream tasks like personalized therapy planning.

\end{abstract}

\section{Introduction}

\begin{figure*}[t] 
\includegraphics[width=\textwidth]{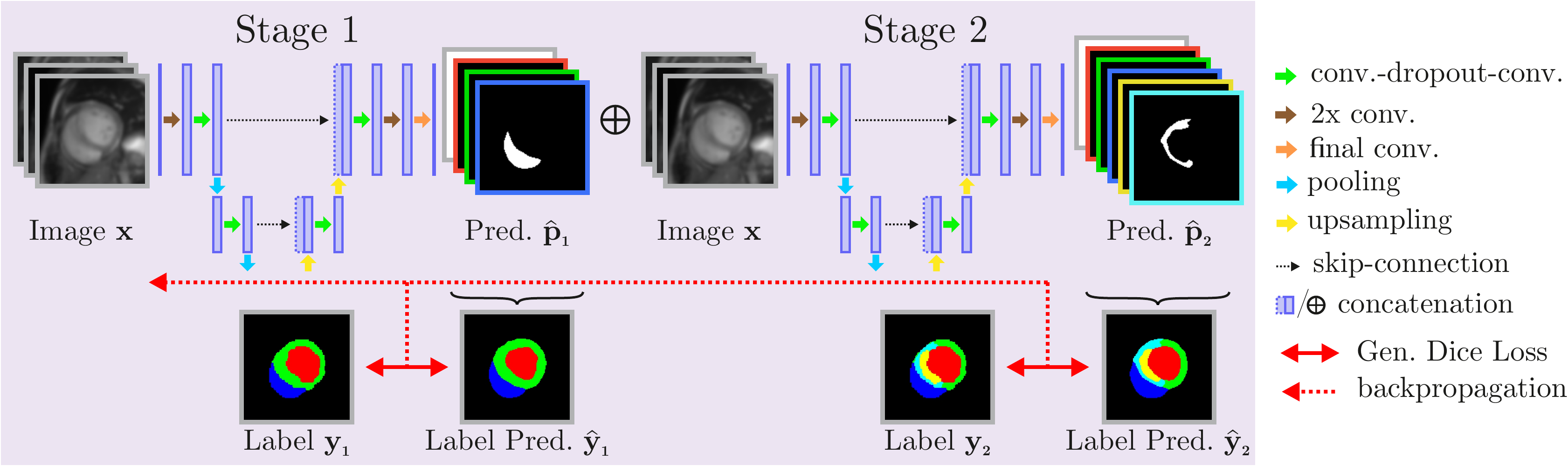}
\caption{
Overview of the proposed method, MS-CaRe-CNN, which is a 2-stage CNN cascade that semantically segments cardiac structures in 3D from multi-sequence data, namely, LGE MR, T2 MR and the end-diastolic phase of bSSFP cine MR.
At Stage~1, given the three MR scans concatenated in channel dimension, MS-CaRe-CNN predicts the left and right ventricle as well as the myocardium which is then concatenated with the original image information in channel dimension.
This initial prediction is then further processed in Stage~2, where the model is trained to additionally distinguish myocardial tissues by separating it into healthy tissue, scar and edema.
}
\label{fig:overview}
\end{figure*}

The leading cause of death worldwide is attributed to cardiovascular diseases of which \gls{mi} is one of the most prevalent~\cite{vaduganathan2022global,salari2023global}.
\gls{mi} occurs when there is a reduction or complete cessation of blood flow in the coronary arteries, which decreases perfusion in the affected myocardial tissue.
This leads to a metabolic undersupply, impairing cardiac function and potentially causing myocardial necrosis, which is a crucial risk factor for developing heart failure.
Even though the global mortality rate directly associated with \gls{mi} has decreased over the last decades~\cite{velagaleti2008long}, the mortality and morbidity rates of heart failure in patients post-\gls{mi} remain high~\cite{juilliere2012heart,lewis2003predictors}.
Diagnosis and treatment planning for patients after acute \gls{mi} relies on accurate assessment of the viability of myocardial tissue and potential tissue damage, crucial indicators for estimating the risk of heart failure.
For example, surgical revascularization may be performed after \gls{mi} aiming to restore normal blood supply to affected yet viable myocardial tissue and to potentially recover functionality~\cite{wroblewski1990evaluation,perin2002assessing}, which improves functional capacity and survival~\cite{van1996magnetic,kim2004viability}.
Furthermore, accurate assessment of infarcted myocardial tissue is essential for determining the risk of adverse cardiovascular events in the future, such as ventricular tachycardia, which can lead to sudden death~\cite{rosenthal1985sudden,hellermann2002heart}.

Different medical imaging techniques with complementary properties can be used in order to allow combining their information and enable a more fine-grained assessment of the myocardial tissue and the surrounding anatomical structures.
So called multi-sequence data aiming to assess myocardial pathology may include, (1) \gls{lge} \gls{mr} imaging, which allows visualization of myocardial scar tissue, (2) \gls{t2} cardiac \gls{mr} imaging, which enables visualization of edema, and (3) \gls{bssfp} cine sequences, that allow distinguishing anatomical boundaries.
On one hand, multi-sequence data is challenging to accurately and efficiently analyze as the individual sequences are often misaligned and need to be registered to each other before being processed.
On the other hand, characterization of myocardial tissue from such data is challenging as the walls are very thin anatomical structures, scarred tissue might form complex patterns and image quality is often limited.
Nowadays, \glspl{cnn} are well-established and widely-adopted methods for automated medical image analysis like the detection of diseases in medical images~\cite{Esteva2017,Feng2022-md}, or image segmentation of the vertebrae~\cite{Payer2020-yv}, or the heart~\cite{Payer2018,chen2020deep}.
Several approaches to semantically segment different cardiac structures while distinguishing between healthy and scarred myocardial tissue from \gls{lge} \gls{mr} scans~\cite{fahmy2018automated,chen2022automatic,xu2022bmanet,thaler2024care}, or from cardiac multi-sequence data~\cite{li2023myops,qiu2023myops,ding2023aligning} have been proposed in literature.
Nevertheless, one central assumption of machine learning algorithms in general is that training and test data are \gls{iid}, i.e. that they are drawn from the same data distribution.
Unfortunately, this assumption is quickly violated, e.g. when test data is obtained from different medical centers or under different conditions like using other scanner models or protocols.
Differences between the training and test dataset are known as domain shift, which results in higher test errors~\cite{Ben-David2006-ck,Torralba2011-gm} or even complete model failure~\cite{AlBadawy2018-gi,Pooch2020-vz} on test data.
By relying on strong data augmentation techniques to diversify the training data as much as possible, data augmentation-based \gls{dg} is a popular group of approaches that aim to address domain shift and improve the performance on domains that are unknown at training-time~\cite{Zhou2023-jk,Wang2023-vz}.

In this work inspired by~\cite{thaler2024care}, we employ \gls{mscarecnn}, which semantically segments myocardial scar tissue and edema among other structures from multi-sequence data, namely, \gls{lge} \gls{mr}, \gls{t2} \gls{mr} and the end-diastolic phase of \gls{bssfp} cine \gls{mr}.
To this end, \gls{mscarecnn} is designed to first predict the anatomical structures of interest, i.e. the left and right ventricle as well as the myocardium overall, before refining these predictions to also consider the viability of the myocardial tissues by separating it into healthy tissue, scar and edema.
\gls{mscarecnn} is a fully 3D \gls{cnn} architecture that consists of two stages and is trained end-to-end.
Our preprocessing pipeline employs strong data augmentation techniques to address domain shift towards domains unknown at training-time by enlarging the diversity of the observable feature representations.
Moreover, training as well as test data is resampled in 3D such that all scans have a consistent and isotropic physical resolution.
Our method is a contribution to the MyoPS++ track of the CARE2024 Challenge\footnote{CARE2024 Challenge website: \url{http://www.zmic.org.cn/care_2024/}, last accessed in September, 2024} and externally evaluated by comparing to other contributions to this challenge.

\section{Method}

We propose \gls{mscarecnn}, a cascading refinement \gls{cnn} which is designed to jointly process multi-sequence scans in 3D.
Our method generates semantic segmentations for the left and right ventricle, healthy and scarred myocardial tissue as well as edema.
\gls{mscarecnn} is closely related to our simultaneous LAScarQS++ track submission named LA-CaRe-CNN~\cite{thaler2024la}, which shows the generality of our cascading refinement \gls{cnn} method.
An overview of \gls{mscarecnn} is provided in Fig.~\ref{fig:overview}.

\boldparagraph{Dataset Grouping:}

While the multi-sequence data encompasses \gls{lge} \gls{mr}, \gls{t2} \gls{mr} and end-diastolic phase of \gls{bssfp} cine \gls{mr} scans, not all sequences are available for all patients.
Moreover, since some labels can only be identified when certain sequences are available, cases with missing sequences also have missing labels.
To account for that, we separated the dataset into three groups based on which sequences and labels are available for each respective case.
Specifically, we refer to the dataset of the first group for which all scans and labels are available as $\datagroupone$, while the dataset of the second group is defined as $\datagrouptwo$ for which the \gls{t2} \gls{mr} scan and the edema label are missing.
Lastly, the dataset of the third group with a missing \gls{t2} \gls{mr} and \gls{bssfp} cine \gls{mr} scan as well as a missing edema and right ventricle segmentation is called $\datagroupthree$.
An overview of available and missing scans and labels per group is provided in Table~\ref{tab:ground_truth}.
For convenience, we will refer to all available image data as the image $\image$, where the individual scans are concatenated in channel dimension and set to zero in case they are missing.
Furthermore, we will refer to the ground truth segmentation as $\groundtruth$, independent of which exact labels are available.

\boldparagraph{Multi-Sequence Cascading Refinement CNN:}

The network architecture of \gls{mscarecnn} is a 2-stage end-to-end trained \gls{cnn} cascade, where two consecutive 3D U-Net-like architectures~\cite{ronneberger2015u} are employed to process the data.
For each iteration during training, we randomly sample the available image data as well as the corresponding ground truth segmentation of one subject of each of the three groups $\datagroupone$, $\datagrouptwo$ and $\datagroupthree$.
After sampling, the images $\imagegroupone$, $\imagegrouptwo$ and $\imagegroupthree$ are provided as independent inputs to Stage~1 of \gls{mscarecnn}, which is designed to generate predictions for the anatomical structures without considering tissue viability.
Namely, the generated predictions only include the left and right ventricle as well as the whole myocardium as three separate labels.
Defining the Stage~1 model as $\modelstageone (\cdot)$ with trainable weights $\modelweightsstageone$ allows formulating the Stage~1 prediction $\predstageone$ for an image $\image$ as:
\begin{equation}
    \predstageone = \modelstageone (\image ; \modelweightsstageone),
\end{equation}
with $\predlettersmall$ referring to the model output before computing any activation function.
After obtaining the Stage~1 prediction for each group as $\predstageonegroupone$, $\predstageonegrouptwo$ and $\predstageonegroupthree$, they are forwarded to the Stage~2 model of \gls{mscarecnn}, which is designed to further process and refine the predictions.
Specifically, Stage~2 generates predictions for all labels, i.e. the left and right ventricle, healthy and scarred myocardial tissue as well as edema.
However, before being processed by the Stage~2 model, the prediction $\predstageone$ of each group is concatenated with the corresponding image $\image$ in channel dimension to allow refinement of the prediction under consideration of the original image information.
Next, we define the Stage~2 model $\modelstagetwo(\cdot)$ with trainable weights $\modelweightsstagetwo$ from which the Stage~2 prediction $\predstagetwo$ can be obtained by computing:
\begin{equation}
    \predstagetwo = \modelstagetwo (\predstageone \oplus \image ; \modelweightsstagetwo),
\end{equation}
where $\oplus$ represents the channel-wise concatenation operator.
The final prediction obtained from Stage~2 for each respective image $\imagegroupone$, $\imagegrouptwo$ and $\imagegroupthree$ is then defined as $\predstagetwogroupone$, $\predstagetwogrouptwo$ and $\predstagetwogroupthree$, respectively.

Our \gls{mscarecnn} is trained fully end-to-end and weight updates are computed via standard backpropagation.
This allows predictions at any stage to contribute to the weight updates of all weights that precede the respective output layer of that stage.
Before computing the loss during training, we apply the activation function to the prediction $\predlettersmall$ in order to obtain the label prediction $\prediction = \text{softmax} ( \predlettersmall )$ for each stage and each group.
Finally, this allows formulating the training objective for \gls{mscarecnn} as:
\begin{equation}
\begin{split}
\lossbase &=
\sum_{g \in \{\text{G1}, \text{G2}, \text{G3}\}}
\Big(
\lossfactorone
\underbrace{
\lossdice(\gtstageone^{g}, \labelpredstageone^{g}; \modelweightsstageone)
}_{\text{update $\modelstageone$}}
+ \lossfactortwo
\underbrace{
\lossdice(\gtstagetwo^{g}, \labelpredstagetwo^{g}; \modelweightsstageone, \modelweightsstagetwo)
}_{\text{update $\modelstageone$ and $\modelstagetwo$}},
\Big)
\end{split}
\label{eq:mlc_loss}
\end{equation}
where the corresponding ground truth segmentation $\groundtruth$ follows the same label definition of Stage~1 and 2, respectively.
Furthermore, please note that the loss is only computed for labels that are available in the respective ground truth.
$\lossdice$ refers to the generalized Dice loss function and the weighting factors $\lossfactorone$ and $\lossfactortwo$ are both defined as 1.

\boldparagraph{Addressing Domain Shift:}

Similar to popular strategies for \gls{dg}~\cite{Zhou2023-jk,Wang2023-vz}, we account for domain shift towards domains that are unknown at training-time by employing strong data augmentation techniques.
Specifically, we address potential differences in local cohorts like the size and morphology of the anatomy of interest by using translation, rotation, scaling and elastic deformation as spatial data augmentation techniques.
Please note that the same spatial augmentation is performed for each sequence that belongs to the same subject when selected for training in order to not destroy their spatial correspondence.
Additionally, intensity augmentation techniques are employed to allow modulation of intensity values to mimic differences in signal-to-noise ratio, intensity ranges and contrast that might be introduced by different scanner models or acquisition protocols.
To achieve this, we globally modify the intensity values during training before modulating them per label by randomly sampling sequence-specific shift and scale parameters.

\begin{table*}
\centering
\caption{
The number of subjects per center as well as the available sequences that constitute the training, validation and test set of the MyoPS++ track of the CARE2024 Challenge.
}
\begin{tabular}{
p{0.2\textwidth}
| >{\centering}p{0.1\textwidth}
| >{\centering}p{0.1\textwidth}
| >{\centering}p{0.1\textwidth}
| >{\centering}p{0.1\textwidth}
| >{\centering}p{0.1\textwidth}
| >{\centering}p{0.1\textwidth}
| >{\centering\arraybackslash}p{0.1\textwidth}
}

Sequences & \multicolumn{1}{c|}{LGE} & \multicolumn{3}{c|}{LGE, T2, bSSFP} & \multicolumn{3}{c}{LGE, bSSFP} \\
Center & A & B & C & D & E & F & G \\
\hline

Training Set & 81 & 50 & 45 & -- & 7 & 9 & 8 \\  
Validation Set & -- & -- & -- & 25 & -- & -- & -- \\
Test Set & -- & -- & -- & 25 & -- & -- & -- \\

\end{tabular}

\label{tab:dataset}
\end{table*}

\section{Experimental Setup}

\subsection{Dataset}

The dataset used in this work is part of the CARE2024 Challenge and provided for the MyoPS++ track.
The MyoPS++ dataset is a multi-sequence dataset which includes \gls{lge} \gls{mr}, \gls{t2} \gls{mr} and \gls{bssfp} cine \gls{mr} at the end-diastolic phase of the cardiac cycle.
Overall, \gls{mr} scans of 250 patients obtained from seven medical centers are available, however, not all sequences exist for all cases.
Specifically, for 145 patients \gls{lge}, \gls{t2} and \gls{bssfp} cine \gls{mr} scans are available, for 24 patients only \gls{lge} and \gls{bssfp} cine \gls{mr} scans exist and finally, for 81 patients only \gls{lge} \gls{mr} scans are available, see Table~\ref{tab:dataset}.
Furthermore, while overall five ground truth labels are defined, not all ground truth labels are available for all scans, see Table~\ref{tab:ground_truth}.
The ground truth labels include the left and right ventricle, healthy and scarred myocardial tissue as well as the edema.
The exact numbers of the training, validation and test set per center for both tracks are provided in Table~\ref{tab:dataset}.
All data was provided after registration using a multivariate mixture model by the organizers~\cite{zhuang2019multivariate}.

\begin{table*}
\centering
\caption{
Detailed information on which centers are part of the three dataset groups ($\datagroupone$, $\datagrouptwo$ and $\datagroupthree$) we used to train MS-CaRe-CNN.
Moreover, we provide information on which sequences and which ground truth labels are available per group and center.
}
\begin{tabular}{
>{\centering}p{0.08\textwidth}
| p{0.1\textwidth}
| >{\centering}p{0.08\textwidth}
| >{\centering}p{0.08\textwidth}
| >{\centering}p{0.08\textwidth}
| >{\centering}p{0.08\textwidth}
| >{\centering}p{0.08\textwidth}
| >{\centering}p{0.08\textwidth}
| >{\centering}p{0.08\textwidth}
| >{\centering\arraybackslash}p{0.08\textwidth}
}

\multirow{2}{*}{Group} & \multirow{2}{*}{Center} & \multicolumn{3}{c|}{Sequences} & \multicolumn{5}{c}{Ground truth labels} \\
& & LGE & T2 & bSSFP & LV & RV & MYO & Scar & Edema \\
\hline

$\datagroupone$ & B, C, D & \checkmark & \checkmark & \checkmark & \checkmark & \checkmark & \checkmark & \checkmark & \checkmark \\
$\datagrouptwo$ & E, F, G & \checkmark & & \checkmark & \checkmark & \checkmark & \checkmark & \checkmark & \\
$\datagroupthree$ & A & \checkmark & & & \checkmark & & \checkmark & \checkmark & \\

\end{tabular}

\label{tab:ground_truth}
\end{table*}

\subsection{Implementation Details}

In our preprocessing pipeline, we resample all training, validation and test data of each group to the same physical resolution of $1.2 \times 1.2 \times 1.2$ mm.
Since the GPU memory requirements are larger at training-time, we compute the center position from the ground truth segmentation of every subject in the training set.
This center position is then used as the center point of the region of interest which we extract for further processing with the size $128 \times 128 \times 128$ voxel.
For subjects in the validation and test set, we treat the center position of the whole scan as the center point for the region of interest and extract an image with an increased size of $192 \times 192 \times 192$ voxel for these cases.

At training-time, we employ spatial and intensity augmentation techniques in 3D~\cite{Payer2018,payer2019integrating} to diversify the available data aiming to improve generalization to unknown domains.
Specifically, spatial augmentation relies on translation ($\pm 20$ voxels), rotation ($\pm 0.35$ radians), isotropic scaling ($[0.8, 1.2]$), anisotropic scaling per dimension ($[0.9, 1.1]$) and elastic deformation (eight grid nodes per dimension and deformation values sampled from $\pm 15$ voxels).
Next, we robustly normalize data by linearly mapping the \nth{10} and \nth{90} percentile of the intensity values to $-1$ and $1$, respectively.
For intensity augmentation, we randomly sample intensity shift ($\pm 0.2$) and scaling factors ($[0.6, 1.4]$) per sequence and subject.
All augmentation parameters are sampled from a uniform distribution within the provided ranges.
Please note that data from the validation and test set is not augmented, however, it is robustly normalized.
Moreover, for validation and test set predictions, we employ a connected component analysis as a post-processing, where components that are disconnected from the largest component in 3D per label as well as for all foreground labels considered as one structure are removed.

Each stage of \gls{mscarecnn} is implemented similarly to a 3D U-Net~\cite{ronneberger2015u}, see Fig.~\ref{fig:overview}.
While the architecture of each stage follows the same structure, their weights are not shared across stages.
The stage-specific architecture consists of a contracting and an expanding path, and has five levels of depth with skip-connections to allow intermediate feature representations to bypass lower levels.
Each level of the contracting and expanding path consists of two convolution layers with a dropout layer in-between~\cite{srivastava2014dropout}, followed by either a max pooling or a linear upsampling layer, respectively.
Moreover, an additional two respectively three convolution layers are used before and after each stage.
Before computing the loss, we employ a separate final convolution layer per stage and group that has the same number of filters as there are labels for the respective stage and group, in order to avoid numerical instabilities during training.
Intermediate convolution layers employ a kernel size of $3 \times 3 \times 3$ and $64$ filters, while the final convolution layer uses a kernel size of $1 \times 1 \times 1$ and as many filters as there are expected outputs.
He initialization~\cite{he2015delving} is used to initialize all convolution kernels before training and the dropout rate is set to $0.1$.
After intermediate convolution layers, leaky ReLU~\cite{maas2013rectifier} is used as activation function with a slope of $0.1$, and softmax activation is used after the final convolution layer when computing the loss.
Furthermore, we use Adam~\cite{kingma2014adam} as the optimizer with a learning rate of $0.0005$ and temporal ensembling~\cite{laine2016temporal} of network weights.
For the final submission, we trained our \gls{cnn} for $80,000$ iterations.
Our final submission employs a 5-fold ensemble of independently trained \glspl{mscarecnn}, which averages the predictions of the individual models.
Training of a single model lasted roughly 38~hours, while inference of the whole ensemble after loading takes about 7~seconds per subject on an NVIDIA GeForce RTX 3090.

\begin{table*}[t]
\centering
\caption{
Quantitative results on the validation set of the MyoPS++ track of the CARE2024 Challenge.
The evaluation includes scores for the Dice Similarity Coefficient (DSC), Sensitivity (SEN), Specificity (SPE) and Precision (PRE) in percent for myocardial scar tissue (Scar) as well as for the joint scar and edema label (Scar\&Edema).
The performance of our final submission (5-fold Ensemble) is compared to the performance of a single model.
The best score per metric is shown in bold.
}
\resizebox{\columnwidth}{!}{%
\begin{tabular}{ l | c | c | c | c | c | c | c | c }

& \multicolumn{4}{c|}{\textbf{Scar}} & \multicolumn{4}{c}{\textbf{Scar\&Edema}} \\
& DSC ($\uparrow$) & SEN ($\uparrow$) & SPE ($\uparrow$) & PRE ($\uparrow$) & DSC ($\uparrow$) & SEN ($\uparrow$) & SPE ($\uparrow$) & PRE ($\uparrow$) \\
\hline


single model & 60.32 & 50.63 & 99.93 & 79.56 & 65.15 & 57.44 & \textbf{99.84} & 79.10 \\
5-fold ensemble & \textbf{62.76} & \textbf{51.83} & \textbf{99.94} & \textbf{84.40} & \textbf{65.36} & \textbf{58.34} & \textbf{99.84} & \textbf{79.46} \\



\end{tabular}
}

\label{tab:quantitative_results}
\end{table*}

\section{Results and Discussion}

\begin{figure*}[t] 
\centering
\includegraphics[width=0.75\textwidth]{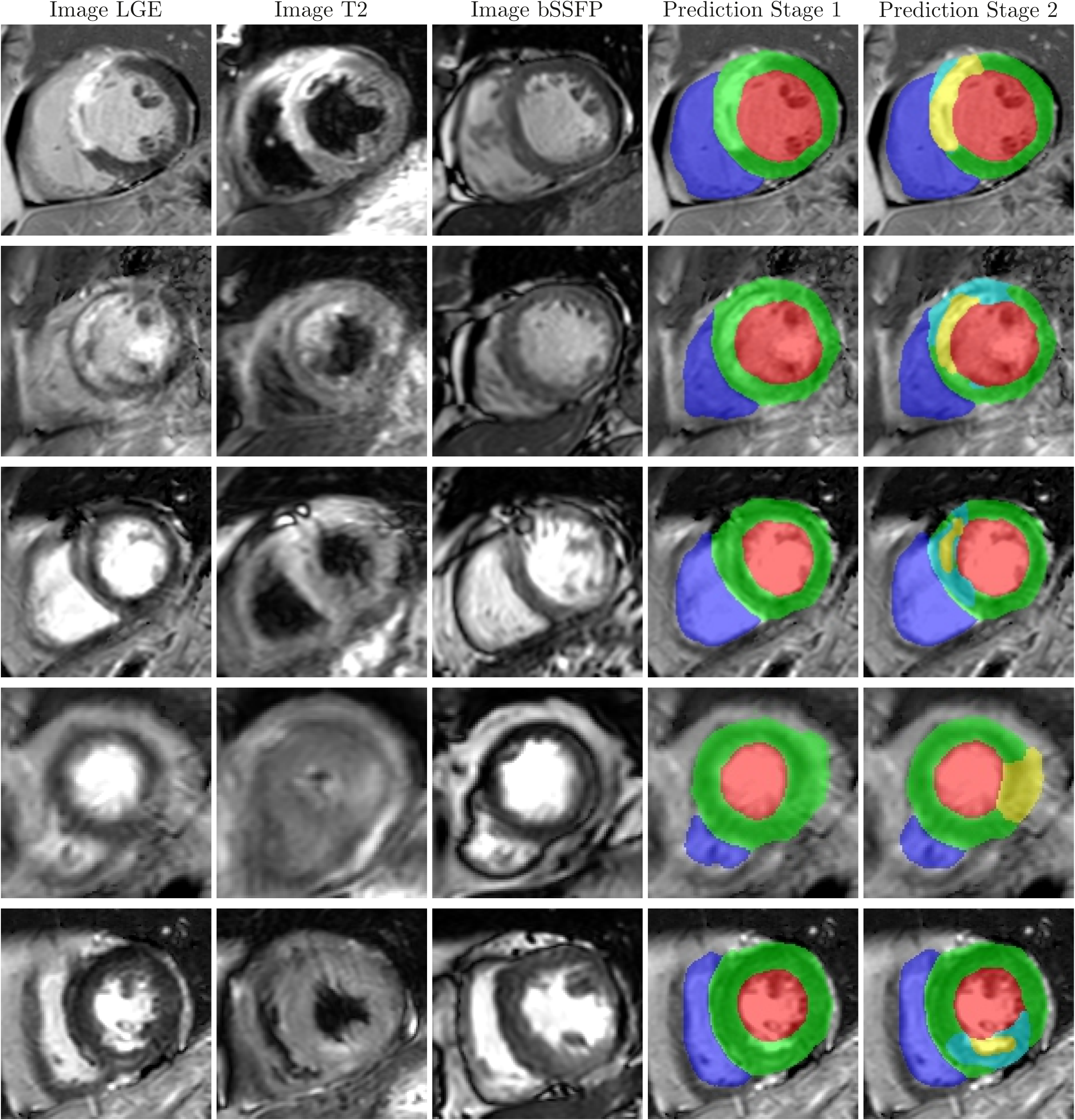}
\caption{
Qualitative results on the validation set of the CARE2024 Challenge MyoPS++ track. 
We show corresponding examples of LGE MR (col. 1), T2 MR (col. 2) and the end-diastolic phase of bSSFP cine MR images (col. 3), as well as predictions of Stage~1 (col. 4) and Stage~2 (col. 5).
In Stage~1, colors represent the left ventricle (red), right ventricle (blue) and myocardium as one label (green).
Colors of Stage~2 refer to the left ventricle (red), right ventricle (blue), healthy myocardial tissue (green), scarred myocardial tissue (yellow) and edema (cyan).
}
\label{fig:qualitative_results}
\end{figure*}

We evaluate the performance of our method on the validation set, for which we use the quantitative scores obtained through the submission system as provided by the organizers of the CARE2024 Challenge.
Specifically, we obtained the scores for four metrics, namely, \gls{dsc}, \gls{sen}, \gls{spe} and \gls{pre} in percent, for two labels.
The two labels are the scar tissue by itself (Scar) as well as the scar and the edema combined into one label (Scar\&Edema).
For the qualitative evaluation on the validation set for which ground truth segmentations are not publicly available, we directly compare the predictions of Stage~1 and Stage~2 to the image information, i.e. \gls{lge} \gls{mr}, \gls{t2} \gls{mr} and \gls{bssfp} cine \gls{mr}, by visual inspection.

The quantitative evaluation results are provided in Table~\ref{tab:quantitative_results}, where we compare the performance of our final submission, a 5-fold ensemble of \glspl{mscarecnn}, to the performance of a single model.
It can be observed that our 5-fold ensemble outperformed the single model on all metrics and both labels, Scar and Scar\&Edema.
This confirms that the 5-fold ensemble, which consists of five independently trained models with different random seeds for initial weights and augmentation parameters, is able to achieve better generalization to previously unknown domains by averaging the individual predictions.

In Fig.~\ref{fig:qualitative_results} we present qualitative results of our method, for which we provide several examples showing the corresponding input images, i.e. \gls{lge} \gls{mr} (col 1), \gls{t2} \gls{mr} (col 2), \gls{bssfp} cine \gls{mr} (col 3) as well as the Stage~1 (col 4) and Stage~2 prediction (col 5) of our model.
For Stage~1, the labels represent the left ventricle (red), right ventricle (blue) and full myocardium \textit{without} distinguishing healthy tissue from scarred tissue or edema (green).
The Stage~2 prediction shows the left ventricle (red), right ventricle (blue), healthy myocardial tissue (green), scar (yellow) and edema (cyan).
Overall, the predictions of the individual labels in Stage~1 and Stage~2 are convincing, however, some inaccuracies especially with the myocardium can be observed.
For example, some oversegmentation of the myocardium (green) in Stage~1 and the edema (cyan) in Stage~2 can be observed in the top left area of the myocardium (col 4-5, row 1, 2).
Moreover, a similar oversegmentation is visible for the myocardium (green) in Stage~1 and myocardial scar tissue (yellow) in Stage~2 in the right area of the myocardium (col 4-5, row 4).
Upon closer inspection of the aligned training data set, we noticed that some of these inaccuracies might be introduced by the multivariate mixture model, which was used by the organizers to register the individual scans to one another and also to combine the individual segmentations of the input images into one joint segmentation~\cite{zhuang2019multivariate}.
It would be interesting to experiment with different registration methods in the future to assess whether more robust segmentation models can be obtained by training on differently registered data.
Nevertheless, the majority of the structures were correctly segmented by our method.

\section{Conclusion}

In this work we presented \gls{mscarecnn}, which is designed to generate semantic segmentations for cardiac structures from multi-sequence data, namely, \gls{lge} \gls{mr}, \gls{t2} \gls{mr} and the end-diastolic phase of \gls{bssfp} cine \gls{mr} scans.
\gls{mscarecnn} is an end-to-end trained 2-stage \gls{cnn} cascade that processes data in 3D.
At Stage~1, our \gls{mscarecnn} generates predictions for the anatomical structures of interest, i.e. the left ventricle, right ventricle and myocardium without considering tissue viability.
These predictions are then concatenated with the original image information in channel dimension and forwarded to Stage~2, which is trained to distinguish between all labels, namely, left and right ventricle, healthy and scarred myocardial tissue as well as edema.
Aiming to address domain shift towards previously unknown domains, we use strong intensity and spatial data augmentation techniques to enhance the diversity of observable feature representations, a common strategy to achieve \gls{dg}.
Our method which we propose as a 5-fold ensemble achieves a \gls{dsc} of $62.31$\% and a \gls{pre} of $82.65$\% when segmenting the myocardial scar tissue, as well as a 
\gls{dsc} of $63.78$\% and a \gls{pre} of $87.69$\% when segmenting the scar tissue and edema as one label -- both are very promising results for such challenging and small structures.
The results of \gls{mscarecnn} demonstrate that semantic segmentations obtained with our method can be used to accurately assess the viability of myocardial tissue by distinguishing healthy tissue from scars and edema, which is valuable information that enables downstream tasks like personalized therapy planning.

\begin{credits}
\subsubsection{\ackname}
This research was funded in whole or in part by the Austrian Science Fund (FWF) 10.55776/PAT1748423 and also by the CardioTwin grant I6540 from the Austrian Science Fund (FWF).
The computational results presented have been achieved in part using the Vienna Scientific Cluster (VSC).

\subsubsection{\discintname}
The authors have no competing interests to declare that are relevant to the content of this article.

\end{credits}
%
%
%
\bibliographystyle{splncs04}
\bibliography{paperpile_martin}

\begin{thebibliography}{10}
\providecommand{\url}[1]{\texttt{#1}}
\providecommand{\urlprefix}{URL }
\providecommand{\doi}[1]{https://doi.org/#1}

\bibitem{AlBadawy2018-gi}
AlBadawy, E.A., Saha, A., Mazurowski, M.A.: Deep learning for segmentation of brain tumors: Impact of cross-institutional training and testing. Medical Physics  \textbf{45}(3),  1150--1158 (2018). \doi{10.1002/mp.12752}

\bibitem{Ben-David2006-ck}
Ben-David, S., Blitzer, J., Crammer, K., Pereira, F.C.: Analysis of representations for domain adaptation. Advances in Neural Information Processing Systems  \textbf{19},  137--144 (2006). \doi{10.7551/mitpress/7503.003.0022}

\bibitem{chen2020deep}
Chen, C., Qin, C., Qiu, H., Tarroni, G., Duan, J., Bai, W., Rueckert, D.: {Deep Learning for Cardiac Image Segmentation: A Review}. Frontiers in Cardiovascular Medicine  \textbf{7}, ~25 (2020)

\bibitem{chen2022automatic}
Chen, Z., Lalande, A., Salomon, M., Decourselle, T., Pommier, T., Qayyum, A., Shi, J., Perrot, G., Couturier, R.: {Automatic Deep Learning-based Myocardial Infarction Segmentation from Delayed Enhancement MRI}. Computerized Medical Imaging and Graphics  \textbf{95},  102014 (2022)

\bibitem{ding2023aligning}
Ding, W., Li, L., Qiu, J., Wang, S., Huang, L., Chen, Y., Yang, S., Zhuang, X.: Aligning multi-sequence cmr towards fully automated myocardial pathology segmentation. IEEE Transactions on Medical Imaging  (2023)

\bibitem{Esteva2017}
Esteva, A., Kuprel, B., Novoa, R.A., Ko, J., Swetter, S.M., Blau, H.M., Thrun, S.: {Dermatologist-level Classification of Skin Cancer with Deep Neural Networks}. Nature  \textbf{542}(7639),  115--118 (2017)

\bibitem{fahmy2018automated}
Fahmy, A.S., Rausch, J., Neisius, U., Chan, R.H., Maron, M.S., Appelbaum, E., Menze, B., Nezafat, R.: {Automated Cardiac MR Scar Quantification in Hypertrophic Cardiomyopathy Using Deep Convolutional Neural Networks}. JACC: Cardiovascular Imaging  \textbf{11}(12),  1917--1918 (2018)

\bibitem{Feng2022-md}
Feng, S., Liu, Q., Patel, A., Bazai, S.U., Jin, C.K., Kim, J.S., Sarrafzadeh, M., Azzollini, D., Yeoh, J., Kim, E., et~al.: {Automated Pneumothorax Triaging in Chest X-rays in the New Zealand Population Using Deep-learning Algorithms}. Journal of Medical Imaging and Radiation Oncology  \textbf{66}(8),  1035--1043 (Dec 2022). \doi{10.1111/1754-9485.13393}

\bibitem{he2015delving}
He, K., Zhang, X., Ren, S., Sun, J.: {Delving Deep into Rectifiers: Surpassing Human-Level Performance on ImageNet Classification}. In: Proceedings of the IEEE International Conference on Computer Vision. pp. 1026--1034 (2015)

\bibitem{hellermann2002heart}
Hellermann, J.P., Jacobsen, S.J., Gersh, B.J., Rodeheffer, R.J., Reeder, G.S., Roger, V.L.: {Heart Failure after Myocardial Infarction: A Review}. The American Journal of Medicine  \textbf{113}(4),  324--330 (2002)

\bibitem{juilliere2012heart}
Juilli{\`e}re, Y., Cambou, J.P., Bataille, V., Mulak, G., Galinier, M., Gibelin, P., Benamer, H., Bouvaist, H., M{\'e}neveau, N., Tabone, X., et~al.: Heart failure in acute myocardial infarction: a comparison between patients with or without heart failure criteria from the fast-mi registry. Revista Espa{\~n}ola de Cardiolog{\'\i}a (English Edition)  \textbf{65}(4),  326--333 (2012)

\bibitem{kim2004viability}
Kim, R.J., Manning, W.J.: {Viability Assessment by Delayed Enhancement Cardiovascular Magnetic Resonance: Will Low-dose Dobutamine Dull the Shine?} Circulation  \textbf{109}(21),  2476--2479 (2004)

\bibitem{kingma2014adam}
Kingma, D.P., Ba, J.L.: {Adam: A Method for Stochastic Optimization}. In: Proceedings of the International Conference on Learning Representations (2015)

\bibitem{laine2016temporal}
Laine, S., Aila, T.: {Temporal Ensembling for Semi-Supervised Learning}. In: Proceedings of the International Conference on Learning Representations (2016)

\bibitem{lewis2003predictors}
Lewis, E.F., Moye, L.A., Rouleau, J.L., Sacks, F.M., Arnold, J.M.O., Warnica, J.W., Flaker, G.C., Braunwald, E., Pfeffer, M.A.: Predictors of late development of heart failure in stable survivors of myocardial infarction: the care study. Journal of the American College of Cardiology  \textbf{42}(8),  1446--1453 (2003)

\bibitem{li2023myops}
Li, L., Wu, F., Wang, S., Luo, X., Mart{\'\i}n-Isla, C., Zhai, S., Zhang, J., Liu, Y., Zhang, Z., Ankenbrand, M.J., et~al.: Myops: A benchmark of myocardial pathology segmentation combining three-sequence cardiac magnetic resonance images. Medical Image Analysis  \textbf{87},  102808 (2023)

\bibitem{maas2013rectifier}
Maas, A.L., Hannun, A.Y., Ng, A.Y.: {Rectifier Nonlinearities Improve Neural Network Acoustic Models}. In: Proceedings of the International Conference on Machine Learning. vol.~30, p.~3. Atlanta, GA (2013)

\bibitem{Payer2018}
Payer, C., {\v{S}}tern, D., Bischof, H., Urschler, M.: {Multi-label Whole Heart Segmentation using CNNs and Anatomical Label Configurations}. In: Statistical Atlases and Computational Models of the Heart. ACDC and MMWHS Challenges. STACOM 2017. Lecture Notes in Computer Science(). vol. 10663, pp. 190--198. Springer (2018). \doi{10.1007/978-3-319-75541-0_20}

\bibitem{payer2019integrating}
Payer, C., {\v{S}}tern, D., Bischof, H., Urschler, M.: {Integrating spatial configuration into heatmap regression based CNNs for landmark localization}. Medical Image Analysis  \textbf{54},  207--219 (2019). \doi{10.1016/j.media.2017.09.003}

\bibitem{Payer2020-yv}
Payer, C., {\v{S}}tern, D., Bischof, H., Urschler, M.: {Coarse to Fine Vertebrae Localization and Segmentation with SpatialConfiguration-Net and U-Net}. In: 15th International Joint Conference on Computer Vision, Imaging and Computer Graphics Theory and Applications ({VISIGRAPP} 2020) - Volume 5: {VISAPP}. pp. 124--133 (2020). \doi{10.5220/0008975201240133}

\bibitem{perin2002assessing}
Perin, E.C., Silva, G.V., Sarmento-Leite, R., Sousa, A.L., Howell, M., Muthupillai, R., Lambert, B., Vaughn, W.K., Flamm, S.D.: {Assessing Myocardial Viability and Infarct Transmurality with Left Ventricular Electromechanical Mapping in Patients with Stable Coronary Artery Disease: Validation by Delayed-Enhancement Magnetic Resonance Imaging}. Circulation  \textbf{106}(8),  957--961 (2002)

\bibitem{Pooch2020-vz}
Pooch, E.H.P., Ballester, P., Barros, R.C.: {Can We Trust Deep Learning Based Diagnosis? The Impact of Domain Shift in Chest Radiograph Classification}. In: Thoracic Image Analysis. {TIA} 2020 {MICCAI} Workshop. pp. 74--83 (2020). \doi{10.1007/978-3-030-62469-9\_7}

\bibitem{qiu2023myops}
Qiu, J., Li, L., Wang, S., Zhang, K., Chen, Y., Yang, S., Zhuang, X.: Myops-net: Myocardial pathology segmentation with flexible combination of multi-sequence cmr images. Medical image analysis  \textbf{84},  102694 (2023)

\bibitem{ronneberger2015u}
Ronneberger, O., Fischer, P., Brox, T.: {U-Net: Convolutional Networks for Biomedical Image Segmentation}. In: Proceedings of the International Conference on Medical Image Computing and Computer-Assisted Intervention. pp. 234--241 (2015)

\bibitem{rosenthal1985sudden}
Rosenthal, M.E., Oseran, D.S., Gang, E., Peter, T.: {Sudden Cardiac Death following Acute Myocardial Infarction}. American Heart Journal  \textbf{109}(4),  865--876 (1985)

\bibitem{salari2023global}
Salari, N., Morddarvanjoghi, F., Abdolmaleki, A., Rasoulpoor, S., Khaleghi, A.A., Hezarkhani, L.A., Shohaimi, S., Mohammadi, M.: The global prevalence of myocardial infarction: a systematic review and meta-analysis. BMC Cardiovascular Disorders  \textbf{23}(1), ~206 (2023)

\bibitem{srivastava2014dropout}
Srivastava, N., Hinton, G., Krizhevsky, A., Sutskever, I., Salakhutdinov, R.: {Dropout: A Simple Way to Prevent Neural Networks from Overfitting}. The Journal of Machine Learning Research  \textbf{15}(1),  1929--1958 (2014)

\bibitem{thaler2024care}
Thaler, F., Gsell, M.A., Plank, G., Urschler, M.: {CaRe-CNN: Cascading Refinement CNN for Myocardial Infarct Segmentation with Microvascular Obstructions}. In: Proceedings of the 19th International Joint Conference on Computer Vision, Imaging and Computer Graphics Theory and Applications (VISIGRAPP 2024) - Volume 3: VISAPP. pp. 53--64 (2024). \doi{10.5220/0012324800003660}

\bibitem{thaler2024la}
Thaler, F., \v{S}tern, D., Plank, G., Urschler, M.: {LA-CaRe-CNN: Cascading Refinement CNN for Left Atrial Scar Segmentation}. In: Toward Real World Medical Image Analysis, MICCAI CARE2024 Challenge (2024)

\bibitem{Torralba2011-gm}
Torralba, A., Efros, A.A.: Unbiased look at dataset bias. In: {IEEE} Conference on Computer Vision and Pattern Recognition ({CVPR}). pp. 1521--1528 (2011). \doi{10.1109/CVPR.2011.5995347}

\bibitem{vaduganathan2022global}
Vaduganathan, M., Mensah, G.A., Turco, J.V., Fuster, V., Roth, G.A.: The global burden of cardiovascular diseases and risk: a compass for future health. Journal of the American College of Cardiology  \textbf{80}(25),  2361--2371 (2022)

\bibitem{velagaleti2008long}
Velagaleti, R.S., Pencina, M.J., Murabito, J.M., Wang, T.J., Parikh, N.I., D'Agostino, R.B., Levy, D., Kannel, W.B., Vasan, R.S.: Long-term trends in the incidence of heart failure after myocardial infarction. Circulation  \textbf{118}(20),  2057--2062 (2008)

\bibitem{van1996magnetic}
Van~der Wall, E., Vliegen, H., De~Roos, A., Bruschke, A.: {Magnetic Resonance Techniques for Assessment of Myocardial Viability}. Journal of Cardiovascular Pharmacology  \textbf{28},  37--44 (1996)

\bibitem{Wang2023-vz}
Wang, J., Lan, C., Liu, C., Ouyang, Y., Qin, T., Lu, W., Chen, Y., Zeng, W., Yu, P.S.: Generalizing to unseen domains: A survey on domain generalization. IEEE Transactions on Knowledge and Data Engineering  \textbf{35}(8),  8052--8072 (2023). \doi{10.1109/TKDE.2022.3178128}

\bibitem{wroblewski1990evaluation}
Wroblewski, L.C., Aisen, A.M., Swanson, S.D., Buda, A.J.: {Evaluation of Myocardial Viability following Ischemic and Reperfusion Injury using Phosphorus 31 Nuclear Magnetic Resonance Spectroscopy in Vivo}. American Heart Journal  \textbf{120}(1),  31--39 (1990)

\bibitem{xu2022bmanet}
Xu, C., Wang, Y., Zhang, D., Han, L., Zhang, Y., Chen, J., Li, S.: {BMAnet: Boundary Mining with Adversarial Learning for Semi-supervised 2D Myocardial Infarction Segmentation}. IEEE Journal of Biomedical and Health Informatics  \textbf{27}(1),  87--96 (2022)

\bibitem{Zhou2023-jk}
Zhou, K., Liu, Z., Qiao, Y., Xiang, T., Loy, C.C.: Domain generalization: A survey. IEEE Transactions on Pattern Analysis and Machine Intelligence  \textbf{45}(4),  4396--4415 (2023). \doi{10.1109/TPAMI.2022.3195549}

\bibitem{zhuang2019multivariate}
Zhuang, X.: Multivariate mixture model for myocardial segmentation combining multi-source images. IEEE Transactions on Pattern Analysis and Machine Intelligence  \textbf{41}(12),  2933--2946 (2019)

\end{thebibliography}

\end{document}